# First Investigation on the Radiation Field of the Gas-Filled Three-Axis Cylindrical Hohlraum

Hang Li, Longfei Jing, Shaoen Jiang[*], Longyu Kuang[†], Huabin Du, Xiayu Zhan, Zhichao Li, Sanwei Li, Liling Li, Jianhua Zheng, Jinhua Zheng, Zhiwei Lin, Lu Zhang, Qiangqiang Wang, Yimeng Yang, Bo Ma, Peng Wang, Dong Yang, Feng Wang, Jiamin Yang, Lin Gao, Haijun Zhang, Juan Zhang, Honglian Wang, Chenggang Ye, Qianqian Gu, Jie Tang, Wei Zhang, Jun Xie, Guanghui Yuan, Zhibing He, Kai Du, Xu Chen, Xiaoxia Huang, Yuancheng Wang, Xuewei Deng, Wei Zhou, Liquan Wang, Caibo Deng, Yongkun Ding, Baohan Zhang

*Laser Fusion Research Center, China Academy of Engineering Physics, P.O.Box 919-986, Mianyang 621900, China*

A novel ignition hohlraum named three-axis cylindrical hohlraum (TACH) is designed for indirect-drive inertial confinement fusion. TACH is a kind of 6 laser entrance holes (LEHs) hohlraum, which is orthogonally jointed of three cylindrical hohlraums. The first experiment on the radiation field of TACH was performed on Shenguang III laser facility. 24 laser beams were elected and injected into 6 LEHs quasi-symmetrically. Total laser energy was about 59 kJ, and the peak radiation temperature reached about 192eV. Radiation temperature and pinhole images in gas-filled hohlraum are largely identical but with minor differences with those in vacuum hohlraum. All laser energy can be totally delivered into hohlraum in 3 ns duration even without filled gas in the Φ1.4mm hohlraum. Plasma filling cannot be obviously suppressed even with 0.5 atm pressure gas in the small hohlraum. Backscattering fractions of vacuum hohlraum and gas-filled hohlraum are both lower than 2%. Experimental study of this new kind of hohlraum can provide guidance for future target design and implosion experiment.

**keyword:** TACH, Shenguang III laser facility, quasi-symmetrical injection, gas-filled

[*]jiangshn@vip.sina.com
[†]kuangly0402@sina.com

## I. INTRODUCTION

In indirect-drive Inertial Confinement Fusion (ICF), laser beams are absorbed by the interior walls of a high Z hohlraum and converted into x-rays, then the x-rays irradiates a low-Z ablator material to bring the fuel in capsule to ignition conditions. Generally, the x-rays radiation is asymmetry and it may lead to asymmetric implosion[1-6]. To achieve the ignition conditions, a convergence ratio of about 30 is necessary in the central hot spot ignition scheme[1,3], so the radiation drive asymmetry should be less than 1%[1], which is the key point for hohlraum design. Up to now, cylindrical hohlraum with 2 LEHs is the main candidate and has been largely studied in the National Ignition Campaign (NIC)[7]. In order to achieve necessary time-varying symmetry in cylindrical hohlraums, multi-cone laser beams are used, and the $P_2$ and $P_4$ asymmetries are controlled by adjusting the power ratio between the inner and outer cones (beam phasing technology)[1,4]. However, the inner cone beams generate a considerable fraction of backscattering[7], and the overlap of multiple cones causes crossed-beam energy transfer[8-10]. In addition, the plasma bubbles generated by outer cone affect the transfer of inner beams. These problems make the beam phasing a very complicated job. In addition, the beam phasing technology strictly depends on simulations, and the plasma of laser plasma interaction (LPI) region is non-local thermodynamic equilibrium, so it is difficult to be accurately calculated[11].

Based on plentiful simulation studies of various hohlraums and existing experimental results on cylindrical hohlraum, a novel hohlraum named three-axis cylindrical hohlraum (TACH) is proposed[12], which is orthogonally jointed of three cylindrical hohlraums, as shown in Fig. 1. The axes of the three cylindrical hohlraums coincide with the X, Y, Z axis of a rectangular coordinate system respectively. TACH is a kind of 6 LEHs hohlraum. Laser beams are injected through every LEH with the same incident angle and in one cone.

A view-factor simulation[13] result shows that the time-varying drive asymmetry of TACH is less than 1.0% in the whole drive pulse period without any supplementary technology. Coupling efficiency of TACH is close to that of 6 LEHs spherical hohlraum[14-16] with corresponding size. Its plasma-filling time is close to that of typical cylindrical ignition hohlraum. Its laser plasma interaction has as low backscattering as the outer cone of the cylindrical ignition hohlraum. Therefore, TACH combines most advantages of various hohlraums and has little predictable risk, providing an important competitive candidate for ignition hohlraum.

The study of TACH was mainly be dependent on numerical simulations and existing experimental results on cylindrical hohlraum before, so experimental studies are critical further.

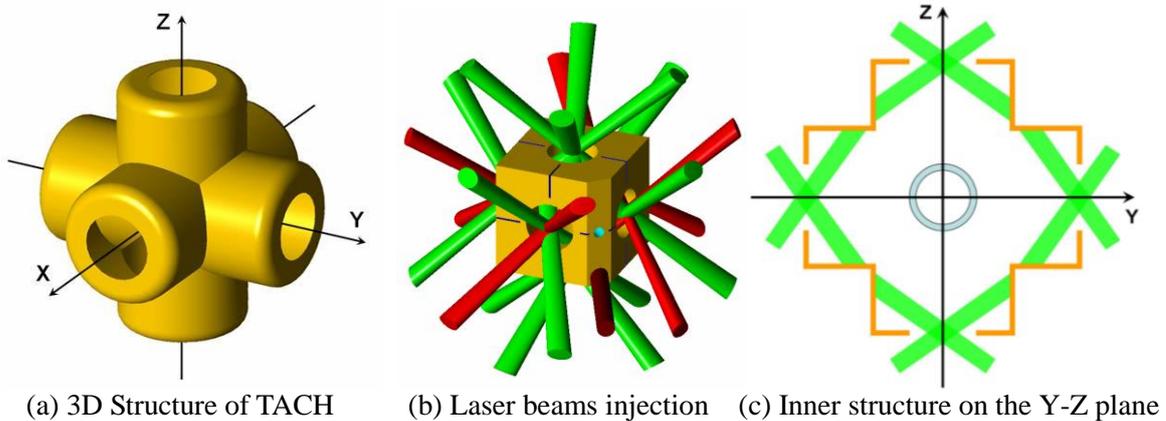

(a) 3D Structure of TACH    (b) Laser beams injection    (c) Inner structure on the Y-Z plane
Figure 1. Schematics of TACH and quasi-symmetrical laser beams injection into 6 LEHs.

In this paper, we report the first experimental study on the radiation field of TACH. Three shots with different density gas-filled hohlraums have been performed on the Shenguang III laser facility[17-18] successfully: vacuum hohlraum, near vacuum low-density gas-filled hohlraum and usual density gas-filled hohlraum. The hohlraum fabrication and laser injection technique of TACH have been developed. There was no capsule in the hohlraum in this experiment. 24 laser beams were injected into 6 LEHs quasi-symmetrically, and every 4 laser beams were injected into one LEH. Radiation temperature, plasma filling and backscatter fraction were studied in this experiment. Experimental study of this new kind of hohlraum can provide guidance for future target design and implosion experiment.

II. EXPERIMENTAL SETUP AND DESIGN

The experiment was performed on the Shenguang III laser facility. The schematic of experimental setup and hohlraum design is shown in Fig. 1. TACH was orthogonally jointed of three ø1.4 mm × 3.2 mm cylindrical hohlraums. Hohlraums with different gas pressure used in expriment were near vacuum low-density 0.1 atm gas-filled hohlraum, usual density 0.5 atm ($n_e$=0.063$n_c$) gas-filled hohlraum and vacuum hohlraum respectively. $C_5H_{12}$ was chosen as the filled gas.

24 laser beams of Shenguang III were chosen, which were injected into 6 LEHs quasi-symmetrically. Every 4 laser beams were injected into one LEH, and laser beams were injected from the center of the corresponding LEH respectively. Incident angle of $\theta_L$ is between 47.88°~50.72° (49.3°±1.42°). Duration of square-shaped main pulse is 3 ns. Trough between pre-pulse and main pulse is 0.5 ns. 0.5 ns square-shaped pre-pulse is used to ablate sealing film of LEH. The power ratio between the pre-pulse and main pulse is 1:10. Total laser energy is about 59 kJ. The laser beams were smoothed by the continuous phase plates (CPPs) with the radius of about 250 μm.

The diagnostic devices included flat-response x-ray detector (F-XRD)[19], M-band x-ray detector (M-XRD), x-ray pinhole camera (XPHC), x-ray framing camera (XFC), full aperture backscatter system (FABS), near backscatter system (NBS), and filter fluoreser hard x-ray detectors (FF). 6 F-XRDs were employed to measure the radiation temperature at different angles, and 6 M-XRD were used to measure M-band fraction in the gold hohlraum. XPHCs and XFC were installed the upper and lower SGIII target chamber to observe the injection of laser beams, which were filtered to record the x-rays at high energy region (~2.5 keV) and low energy region (~0.8 keV). FABS and NBS were employed to measure the backscatter fraction of the incident laser due to the stimulated Brillouin scattering (SBS) and the stimulated Raman scattering (SRS). FF were employed to measure the hot-electron fraction in the hohlraum.

## III. EXPERIMENTAL RESULT AND DISCUSSION

Different field of view was measured by every XRD at different angle of upper 16°, upper 42°, upper 64°, lower 0°, lower 20°, and lower 42°. Different field of view means faced LEH, opposite LEH, laser spots, readiation regions and plasma filling situation are all different. The complex field of view and radiation temperature can be calculated by the three-dimensional view-factor code IRAD3D. But because plasma filling and some other effects have not been taken into consideration in IRAD3D, radiation flux caculated by IRAD3D is inaccurate especially when the viewing angle is small and plasma filling dominates.

As shown in FIG. 2, radiation temperature varies with the different viewing angle. Radiation flux at angle of 0° is the lowest, because there is hardly any hohlraum inner wall in the field of view. At angle of upper 64°, it views small area of laser spots, and its temperature is lower than other angles except 0°. At angle of upper 16° or lower 20°, it views large area of laser spots and also large area of the opposite LEH, so its temperature is a little higher than that at angle of upper 64°. Meanwhile, at angle of upper 16°, it views even larger area of opposite LEH than lower 20°, so its temperature is lower. At angle of upper 42° and lower 42°, it views large area of laser spots and small area of opposite LEH, so its temperature is the highest.

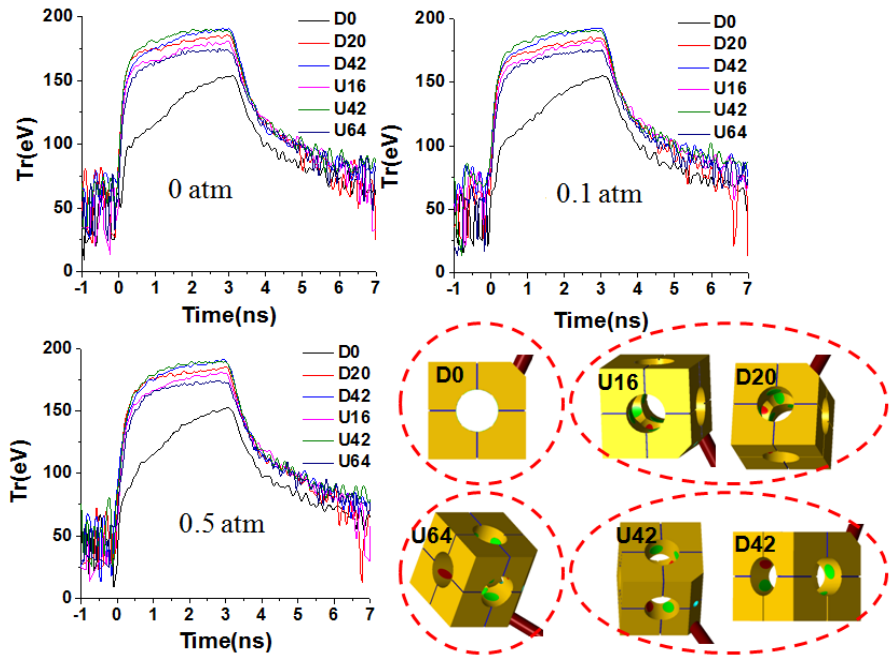

Figure 2. Temporal behavior of radiation temperature at different viewing angles.

FIG. 3 shows radiation temperature at different gas densities from lower 0°, upper 16° and upper 42°. Peak radiation temperature is similar in different shots for a certain angle, which means the final injection of laser beams is similar. However, the measured temporal behavior of radiation temperature is a little different. Radiation temperature of gas-filled hohlraum has a longer rise time since the plasma filling has been slowed down by the filled gas, which is also proved by simulation results.

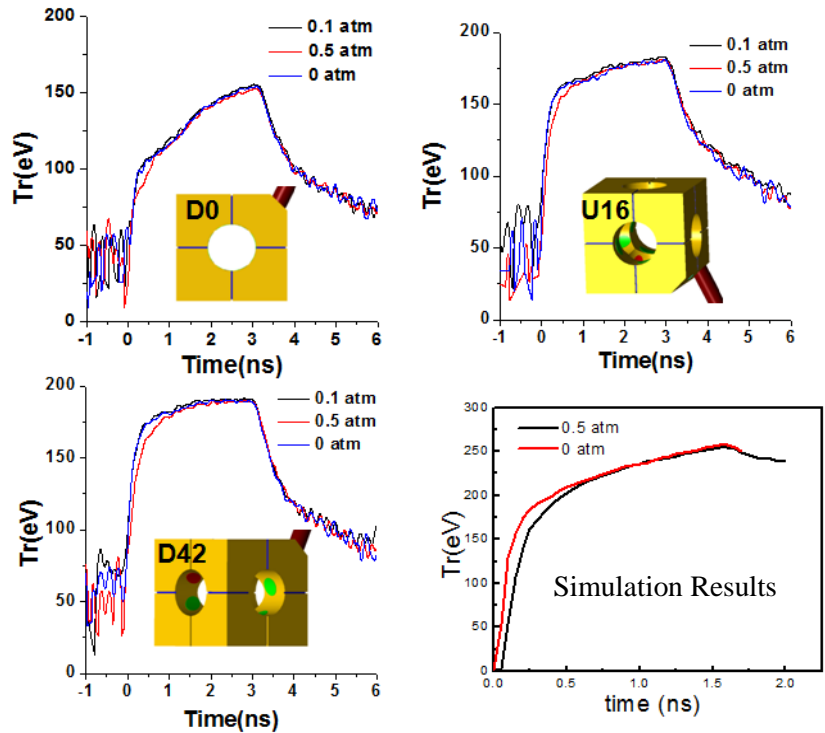

Figure 3. Temporal behavior of radiation temperature at different gas densities.

As shown in FIG. 4, firstly the temporal behavior of radiation temperature of TACH is similar with that of a big traditional cylindrical hohlraum without LEH, which cannot be filled up by plasma and rises slowly in the late period. Normalized radiation temperature of the two kinds of hohlraum at angle of up 42° is almost the same. Secondly, if the plasma filled up the hohlraum in the laser duration and laser directly radiate the filled-up gold plasma, the radiation temperature should rise again by a bigger slope in the late period. But it rises slowly by previous slope, thus the TACH has not been filled up even in the vacuum condition. Meanwhile, the jet will not directly affect the capsule. So in future vacuum hohlraum can be used to study TACH and hohlraum fabrication will be greatly simplified.

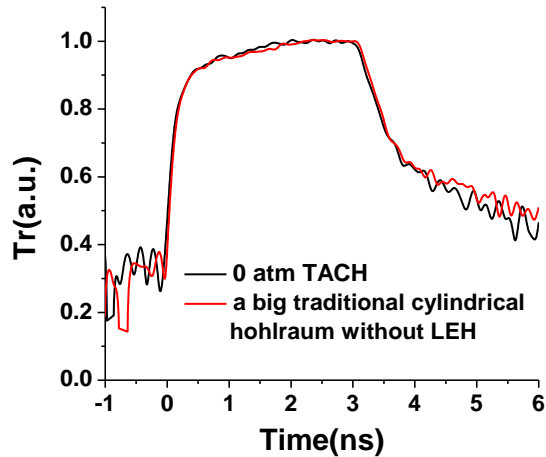

Figure 4. Normalized temporal behavior of radiation temperature of TACH at angle of upper 42° is similar with that of a big traditional cylindrical hohlraum without LEH, which cannot be filled up by plasma and rises slowly in the late period.

Meanwhile, the experimental radiation temperature with that simulated by the IRAD3D are compared, which are shown in FIG. 5. At the angle of upper 16°, lower 20°, upper 42° and lower 42°, the experimental result is consistent with the simulation, and the deviation is less than 3%, which means the IRAD3D can be used to study the multi-holes hohlraum at most angles. However, the deviation is a little large at angle of lower 0° and upper 64°, because IRAD3D has not taken plasma filling into consideration. Therefore, IRAD3D is not suitable for analyze the field of view when the viewing angle is too large or too small.

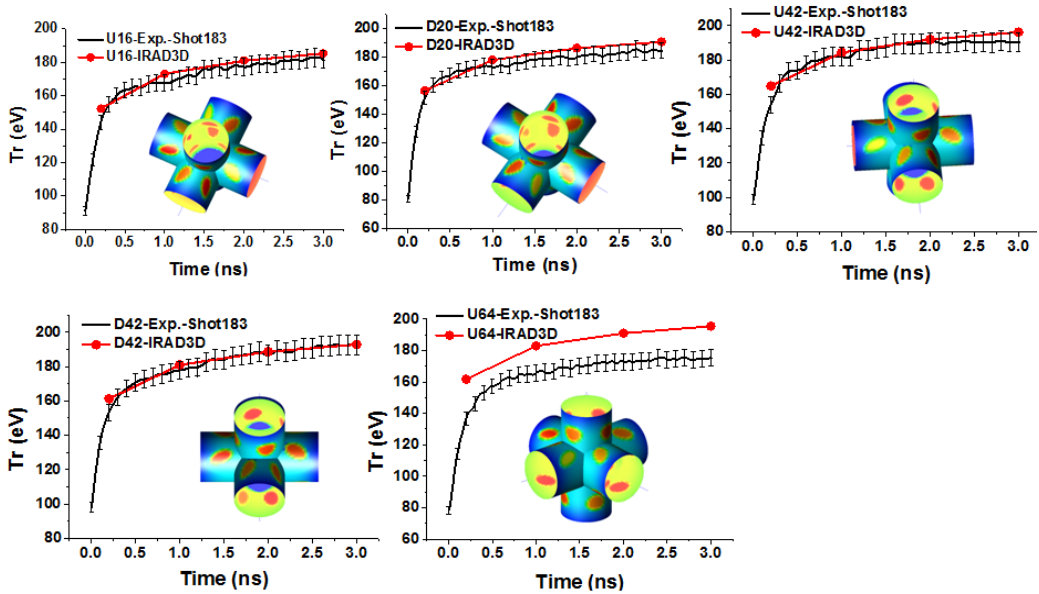

Figure 5. Comparison of experimental result and that simulated by using IRAD3D.

As shown in FIG. 6, the pinhole images are basically consistent between upper and lower XPHC in all shots. In images, brightness indicates x-ray intensity, while x-ray intensity roughly indicates the gold plasma density squared. In the gas-filled hohlraum, a dark cross can be seen, and it is obvious when the gas pressure is higher. The dark cross may be the low-Z gas which hardly emits x-ray and mainly is used to suppress high-Z plasma convergence. While in the vacuum hohlraum, there is a bright cross, which may be caused by convergence of gold plasma bubbles. The pinhole image of the low energy region is not so clear, but the result is similar. Therefore, plasma filling cannot be obviously suppressed even by 0.5 atm gas. Radiation temperature and pinhole images in gas-filled hohlraum is largely identical but with minor differences with those in vacuum hohlraum.

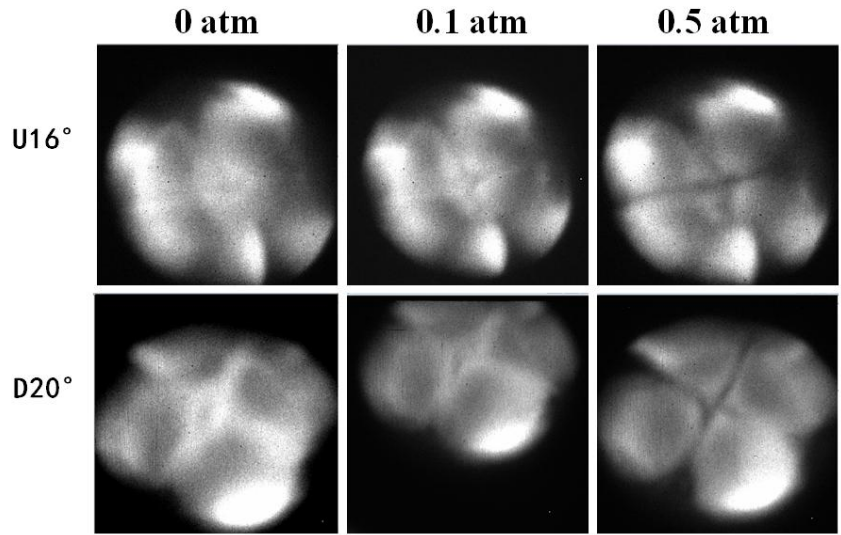

Figure 6. The upper and lower pinhole images of high energy region

Measured backscatter fractions (SRS, SBS and SRS+SBS) from laser beam A4S3 and A4N1 respectively are shown in FIG. 7. Total backscatter fractions in all shots are all lower than 2%. Total backscatter fraction is slightly higher in the gas-filled hohlraum, which is mainly caused by SRS.

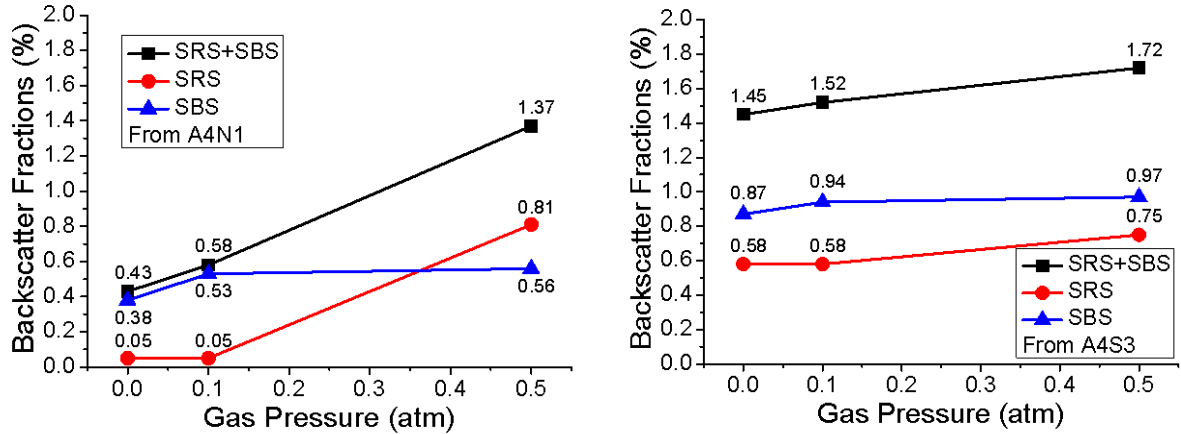
Figure 7. Measured backscatter fractions: SRS, SBS and SRS+SBS from A4N1 and A4S3.

The FF signals for different gas-filled density are all very weak, while signal in the vacuum hohlraum is even weaker.

IV. CONCLUSIONS

In summary, the basic method of TACH fabrication and technique of 24 laser beams injecting quasi-symmetrically into 6 LEHs on Shenguang III have been established. Every 4 laser beams were injected into one LEH with the approximate incident angle of 49.3°±1.42°. Total laser energy was about 59 kJ, and the peak of radiation temperature reached about 192eV. All laser energy can be totally delivered into hohlraum in 3 ns duration even without filled gas in the Φ1.4mm hohlraum. Radiation temperature measured by XRD accords with that simulated by the view-factor code IRAD3D, which proves that IRAD3D can be used to study multi-holes hohlraum effectively. Plasma filling cannot be obviously suppressed even by 0.5 atm gas. Radiation temperature and pinhole images in gas-filled hohlraum are largely identical but with minor differences with those in vacuum hohlraum. Backscattering fractions of vacuum hohlraum and gas-filled hohlraum are both lower than 2%. Total backscatter fraction is slightly higher in the gas-filled hohlraum, which is mainly caused by SRS. Experimental study of this new kind of hohlraum can provide guidance for future target design and implosion experiment.

ACKNOWLEDGMENTS

The authors would like to thank the target fabrication staffs, the laser operation staffs and administrative staffs for their hard work. This work was performed under the auspices of the National Nature Science Fund of China (Grant No. 11475154, 11435011, 11775204, 11505170, 11405160 and 11305160).